\documentclass[a4paper,11pt]{article}
\usepackage{jinstpub} 
\usepackage{lineno}

\usepackage{diagbox}
\usepackage{color}
\usepackage{soul}

\title{\boldmath Noise Filtering Algorithm Based on Graph Neural Network for STCF Drift Chamber}

\author[a]{Xiaoqian Jia,}
\author[a]{Xiaoshuai Qin,}
\author[a]{Teng Li,}
\author[a]{Xueyao Zhang,}
\author[a]{Xiaoqian Hu,}
\author[a]{Shuangbing Song,}
\author[b]{Hang Zhou,}
\author[c]{Xiaocong Ai,}
\author[d]{Jin Zhang,}
\author[a,1]{and Xingtao Huang\note{Corresponding author.}}

\affiliation[a]{Key Laboratory of Particle Physics and Particle Irradiation (MOE), Institute of Frontier and Interdisciplinary Science, Shandong University, Qingdao, Shandong, 266327, China}
\affiliation[b]{Department of Modern Physics, University of Science and Technology of China, Hefei , Anhui, 230026, China}
\affiliation[c]{School of Physics, Zhengzhou University, Zhengzhou, 450001, China}
\affiliation[d]{School of Science, Shenzhen Campus of Sun Yat-sen University, Shenzhen, 518107, China}
\emailAdd{huangxt@sdu.edu.cn}

\abstract{
The super $\tau$-charm facility (STCF) is a next-generation electron-positron collider with high luminosity proposed in China. The higher luminosity leads to increased background level, posing significant challenges for track reconstruction of charged particles. Particularly in the low transverse momentum region, the current track reconstruction algorithm is notably affected by background, resulting in suboptimal reconstruction efficiency and a high fake rate. To address this challenge, we propose a Graph Neural Network (GNN)-based noise filtering algorithm (GNF Algorithm) as a preprocessing step for the track reconstruction. The GNF Algorithm introduces a novel method to convert detector data into graphs and applies a tiered threshold strategy to map GNN-based edge classification results onto signal-noise separation. The study based on Monte Carlo (MC) data shows that with the implementation of the GNF Algorithm, the reconstruction efficiency with the standard background is comparable to the case without background, while the fake rate is significantly reduced. Thus, GNF Algorithm provides essential support for the STCF tracking software.

}


\begin{document}
\maketitle
\flushbottom

\section{Introduction}
\label{sec:intro}

The super $\tau$-charm facility (STCF)~\cite{stcf} is a new generation of electron-positron collider operating in the $\tau$-charm energy region in China, designed with a center-of-mass energy ranging from 2 to 7 GeV and a peak luminosity above $0.5 \times 10^{35} \, \text{cm}^{-2} \text{s}^{-1}$ at $\sqrt{s}=4$ GeV. 
The STCF, currently in the technical design phase, aims to establish a unique platform for the precision frontier of high energy physics. It will address an extensive range of physics topics, including Quantum Chromodynamics (QCD) tests, hadron spectroscopy, CP Violation of mesons and baryons, precise tests of the electroweak sector of the Standard Model (SM), and searches for new physics beyond the SM.

\begin{figure}[htbp]
\centering
\includegraphics[width=.7\textwidth]{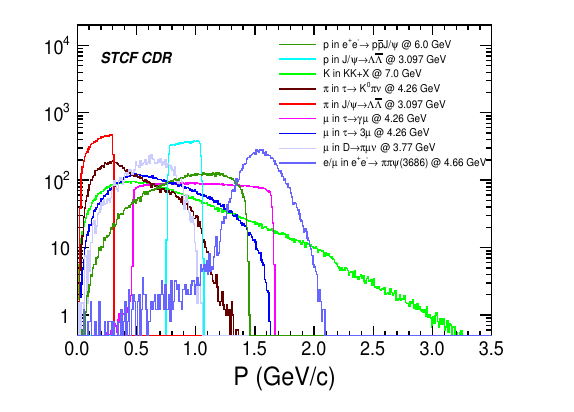}
\caption{Momentum distributions of charged particles from various benchmark physics processes in STCF. This figure is adapted from ref.~\cite{stcf}. \label{fig:cdr}}
\end{figure}

In the high background conditions induced by the high luminosity, it is crucial to detect charged final-state particles with excellent tracking efficiency and momentum resolution. The main sources of background are the luminosity-related background, such as radiative Bhabha scattering and two-photon processes, and beam-related background, including the Touschek and beam-gas effects~\cite{bkg}. The final-state particles, including $e$, $\mu$, $\pi$, $K$ and $p$, have a large momentum coverage range, where most particles have momenta less than 2.0 GeV/c (see figure~\ref{fig:cdr}). In addition, there are a considerable number of particles with momenta below 0.6 GeV/c. Therefore, the track reconstruction algorithm must be able to efficiently reconstruct charged particles across a broad momentum range. The STCF’s baseline track reconstruction algorithm relies on a global track finding algorithm based on Hough transform~\cite{hough} and a track fitting algorithm provided by GENFIT2~\cite{genfit}. This method demonstrates excellent reconstruction efficiency and parameter resolution in high-momentum regions by effectively suppressing background interference. However, its performance deteriorates significantly in low-momentum regimes (particularly below 0.6 GeV/c), where background becomes a limitation. Moreover, the current Hough-based method’s high fake rate leads to increased computational resource consumption (e.g., memory allocation) and much more reconstruction time. To address these challenges, it is essential and urgent to implement an efficient noise filtering preprocessing stage for the Hough-based method.

Inspired by the breakthroughs in the TrackML Particle Tracking Challenge~\cite{trackML1} and the HEP.TrkX project~\cite{trkx1,trkx2}, where GNNs~\cite{GNN} have emerged as promising technologies for addressing tracking challenges in sophisticated silicon detectors~\cite{gnn1,gnn2}, we explore GNNs for the drift chamber track reconstruction. 
Compared to silicon detectors that directly provide high-precision hit positions~\cite{cepc}, drift chambers only provide discrete wire locations and drift time measurements. Therefore, the GNN architecture applied to drift chambers must be capable of comprehensively extracting relationships from lower-resolution raw data to reconstruct true trajectories. We previously developed a track reconstruction algorithm for BESIII~\cite{bes} based on GNN and achieved promising results~\cite{jia}. However, the drift chamber geometry of STCF is different from that of BESIII, and the background level is about an order of magnitude higher than BESIII, so it is necessary to develop a high-performance algorithm to meet the stringent tracking requirements of STCF.

The paper is organized as follows: Section~\ref{sec:2} introduces the STCF drift chamber and a global track finding algorithm based on the Hough transform. Section~\ref{sec:3} describes the design and implementation of the GNN-based noise filtering algorithm (GNF Algorithm). Section~\ref{sec:4} evaluates the performance of the GNF Algorithm under varying background conditions using simulated $J/\psi \rightarrow \pi^0\pi^+\pi^- \rightarrow \gamma\gamma\pi^+\pi^-$ events. Finally, section~\ref{sec:5} gives the summary and outlook.

\begin{figure}[htbp]
\centering
\includegraphics[width=.7\textwidth]{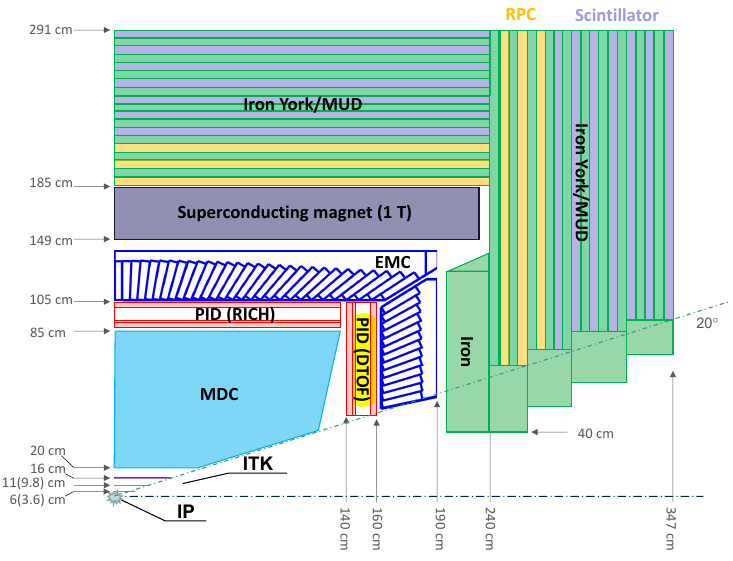}
\caption{Schematic layout of the STCF detector.\label{fig:1}}
\end{figure}

\section{The STCF tracking system and existing tracking strategies}
\label{sec:2}

The STCF detector, shown in figure~\ref{fig:1}, consists of the following subsystems arranged from the interaction point outward: a tracking system, a particle identification system (a Ring Imaging Cherenkov detector in the barrel and a DIRC-like time-of-flight detector in the endcap), a homogeneous Electro-magnetic Calorimeter, a superconducting solenoid, and a Muon Detector. The tracking system consists of an inner tracker (ITK) and a Main Drift Chamber (MDC), combining the ITK's high-precision vertex resolution with the MDC's large-volume trajectory coverage to reconstruct three-dimensional (3D) charged particle trajectories.
The ITK, closest to the beam pipe and covering a polar angle range of $20^{\circ}$ to $160^{\circ}$, is designed to achieve high tracking efficiency for charged particles with very low momentum and improve vertex reconstruction. During the conceptual design phase, the baseline scheme for the ITK detector is three layers of light-material $\mu$RWELL-based gaseous detectors ~\cite{itk}, while a silicon pixel detector serves as an alternative option.

The MDC is the main part of the tracking system of the STCF detector, which combines the ITK to reconstruct the tracks of charged particles and provides the energy loss information to facilitate particle identification. The MDC detector consists of 48 layers of drift cells, with every six layers forming a superlayer. Each superlayer maintains an identical number of cells, though the cell size increases progressively from inner to outer layers, approximately $9.8 \times 9.8 \,\mathrm{mm}^2$ to $12.5 \times 12.5 \,\mathrm{mm}^2$ in the first superlayer and $13.3 \times 13.3 \,\mathrm{mm}^2$ to $14.5 \times 14.5 \,\mathrm{mm}^2$ in the outermost superlayer. The eight superlayers are arranged in an "AUVAUVAA" pattern, with axial ("A") layers aligned parallel to the beam direction and stereo ("U" or "V") layers aligned at specific angles relative to the beam. The "U" and "V" layers have opposing stereo angles to ensure 3D measurement. According to physics requirements, the tracking system needs to achieve track reconstruction efficiencies exceeding 99\% for charged particles in the high transverse momentum region ($p_{T}$ > 300 MeV/c) while maintaining 90\% efficiency at $p_{T}$ = 100 MeV/c. Additionally, it should provide a momentum resolution of $\sigma_{p_{T}} \slash {p_{T}}$ < 0.5\% for tracks with $p_{T}$ = 1 GeV/c~\cite{stcf}.

A Hough transform-based global tracking algorithm is being implemented within the STCF offline software framework OSCAR~\cite{oscar1,oscar2} as a basic approach of the current track finding scheme. The Hough transform is widely used as an image processing technique for detecting geometric shapes, especially for straight lines or circles, by mapping points in the real space to a parameter space. 
Ideally, charged particle trajectories follow helical paths along the $z$-axis of the MDC. Given that the track projection onto the $x-y$ plane is approximately circular, this track finding process is divided into two steps: searching for circular tracks in the transverse $x-y$ plane, followed by the reconstruction of helical trajectories in 3D space. The two-dimensional (2D) track finding starts with a conformal mapping, where each hit from ITK and MDC axial layers is mapped to a point in the conformal plane, which is subsequently projected onto the parameter plane through the Hough transform. Peak searching is performed in the parameter plane to find candidate tracks, and the hit residuals are used to select the hits corresponding to the tracks. After identifying the 2D circular tracks, the algorithm uses their parameters to select the stereo wire hits close to the tracks and performs 3D track finding and fitting. Finally, these candidate tracks are fitted using the Deterministic Annealing Filter (DAF) algorithm provided by GENFIT2 to obtain the track parameters. Recently, a combination of the Hough Transform and the Combinatorial Kalman Filter algorithm in A Common Tracking Software (ACTS)~\cite{acts,actsTrack1} to enhance track finding for long-lived particles at STCF was investigated~\cite{actsLong}.

\section{Design and implementation of the GNF Algorithm}
\label{sec:3}

This section describes the transformation of raw hit data into graph structures, where edge pre-selection faces the critical challenge of ensuring complete inclusion of true track segments while controlling graph size. We subsequently present an edge-classifying GNN architecture: An edge is considered true if its two nodes originate from the same track; otherwise, it is classified as fake. The GNN output is a weight for each edge, representing the probability of this edge being a true edge. Finally, we use these edge weights to distinguish signal from noise according to a hierarchical threshold strategy.

\begin{figure}[htbp]
\centering
\includegraphics[width=.9\textwidth]{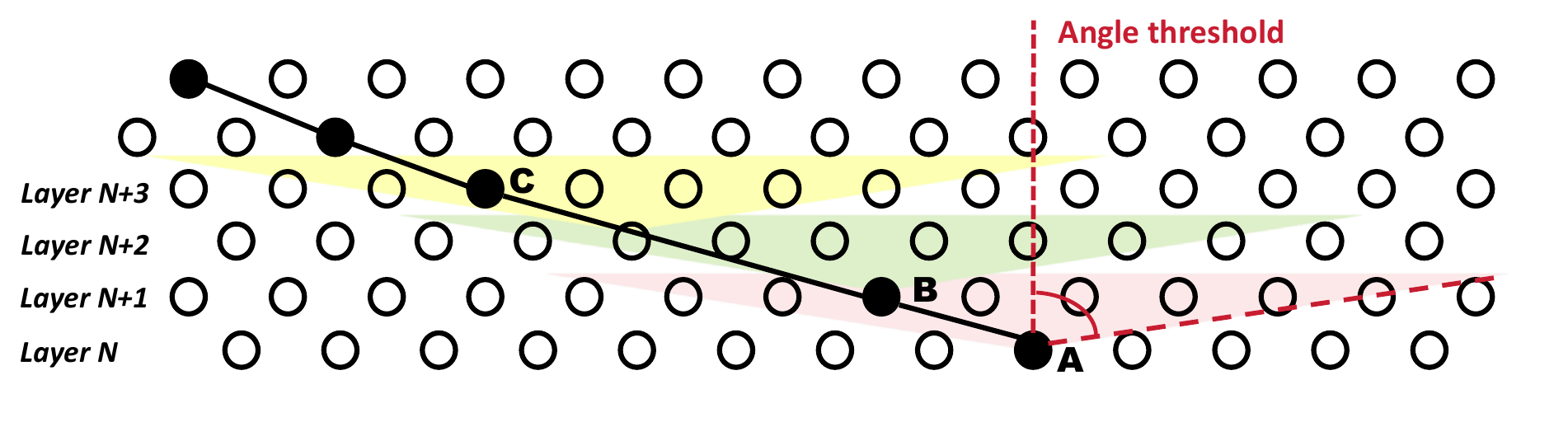}
\caption{Candidate regions and hit connection method. (Colored triangular regions represent next-layer candidate regions for certain sense wires.)\label{fig:3}}
\end{figure}

\subsection{Graph construction}

High-statistics single-particle event samples ($10^5$ events/particle) are generated for each type of charged particle ($e$, $\mu$, $\pi$, $K$ and $p$) to study their motion patterns, with momenta spanning from 0.05 GeV/c to 3.5 GeV/c. The deflection angle of particle trajectories between two adjacent sense layers in the $x-y$ plane is influenced by both the transverse momentum of particles and the geometry of sense wires. Based on the observed maximum deflection angle between adjacent layers, an angle threshold was determined for each of the first 47 layers. We defined the candidate regions using the angle thresholds, as shown in the red area in figure~\ref{fig:3}. Most layers use a threshold of 1.24 rad, while at the boundary of the "U" and "V" superlayers, the threshold is set to 1.5 rad.

At the graph construction stage, each node in the graph corresponds to a hit, and the assignment of edges is based on the candidate regions:

1. An edge connects hit \textit{A} ($Layer_{\textit{N}}$ ) to hit  \textit{B} ($Layer_{\textit{N+1}}$ ) when \textit{B} resides in \textit{A}'s candidate region. 

2. If there is no hit in $Layer_{\textit{N+2}}$, but a hit \textit{C} in $Layer_{\textit{N+3}}$ lies within the candidate regions of any neighboring hit to \textit{B} in $Layer_{\textit{N+2}}$, \textit{B} and \textit{C} are directly connected (figure~\ref{fig:3}), preserving track continuity.

3. Edges are also built between adjacent hits within the same layer.

\begin{figure}[htbp]
\centering
\includegraphics[width=1.\textwidth]{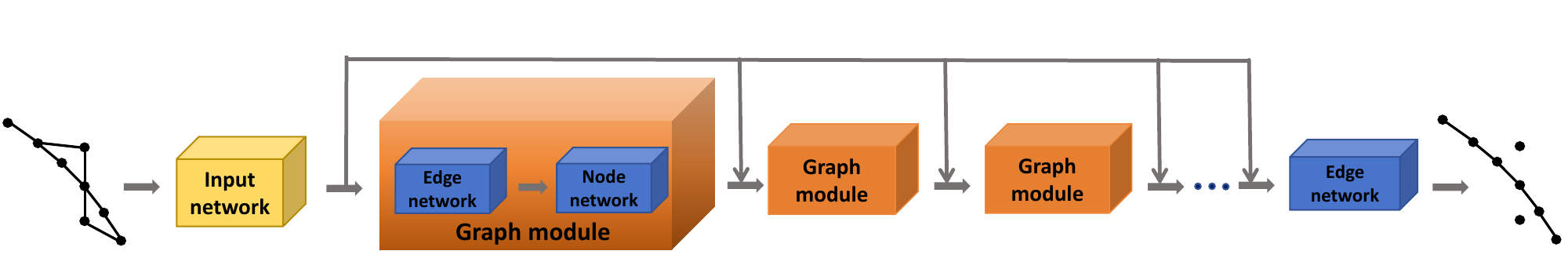}
\caption{Architecture of Graph Neural Network.\label{fig:4}}
\end{figure}

\subsection{Hit classification}
The GNN model is implemented within the PyTorch Geometric framework, which offers modular and high-performance primitives for graph learning. The GNN node features comprise the following physical observables: (1) the position of the center point of the fired sense wires, which has been tested to be better represented by ($r$, $\phi$) coordinates than ($x$, $y$) coordinates; (2) Charge collection measured by the Analog-to-Digital Converter (ADC), corresponding to the energy loss of charged particles in the drift cell;
(3) Time information recorded by the Time-to-Digital Converter (TDC), initiated at the particle's passage through the chamber and terminated at electron collection on the anodic wire.

The input features are first encoded into latent representations through a two-layer fully connected neural network before entering the graph module. The graph module contains two core components: an edge network computes connection probabilities between node pairs by processing their concatenated features through a multi-layer perceptron (MLP) with layer normalization. Edge weights are calculated via the sigmoid activation function. A node network is an MLP with layer normalization. It updates node representations by aggregating weighted neighbor features from both input and output edges, followed by a non-linear projection (Tanh activation) and residual connections to preserve original features.
By iterating this graph module multiple times and concatenating the original features with hidden features after each iteration, multi-hop information propagation is achieved, enabling nodes to progressively integrate features from distant neighbors through consecutive edge-weighted aggregations while optimizing gradient flow. Tests show that eight iterations achieved optimal results. Finally, the edge weights are updated once via the edge network to produce the model's output. The model architecture is shown in figure~\ref{fig:4}. The output edge weight represents the probability of connection between two nodes. The nodes connected to edges that exceed the threshold are considered signals, while the opposite is noise. If a node reaches different conclusions on different edges, we prioritize considering it as a signal. The network employs the Adam optimizer~\cite{adam} with a binary cross-entropy loss function.

\section{Performance of the GNF Algorithm}
\label{sec:4}
$J/\psi \rightarrow \pi^0\pi^+\pi^- \rightarrow \gamma\gamma\pi^+\pi^-$ events are used for the noise filtering performance studies. Its final-state particles reach transverse momenta up to 1.5 GeV/c. For higher momentum, the track reconstruction efficiency is high and almost unaffected by noise, which is not within the scope of this study.

\subsection{Data samples}
\label{sec:4-1}
\begin{figure}[htbp]
\centering
\includegraphics[width=.32\textwidth]{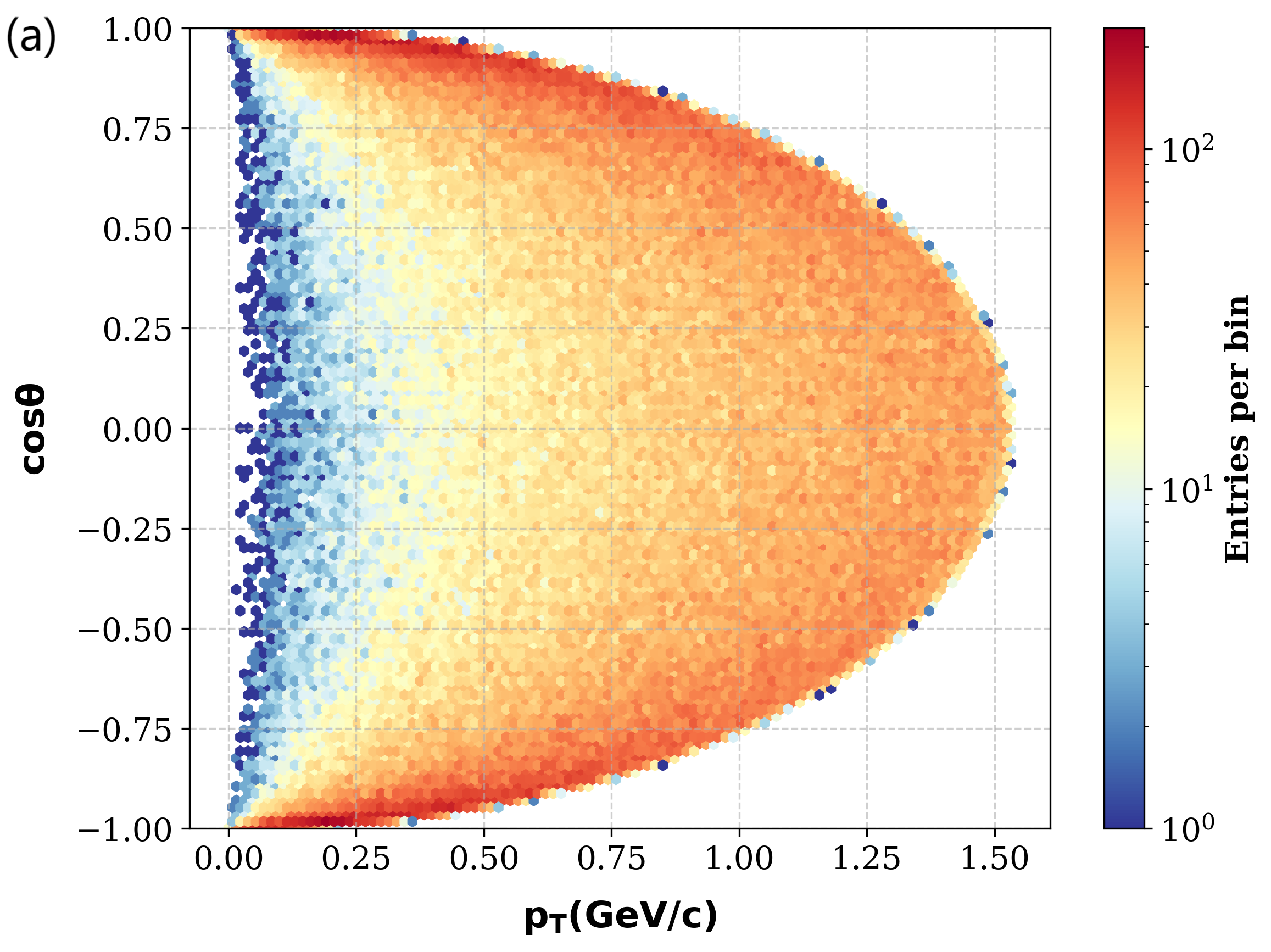}%
\hspace{0.01\textwidth}
\includegraphics[width=.32\textwidth]{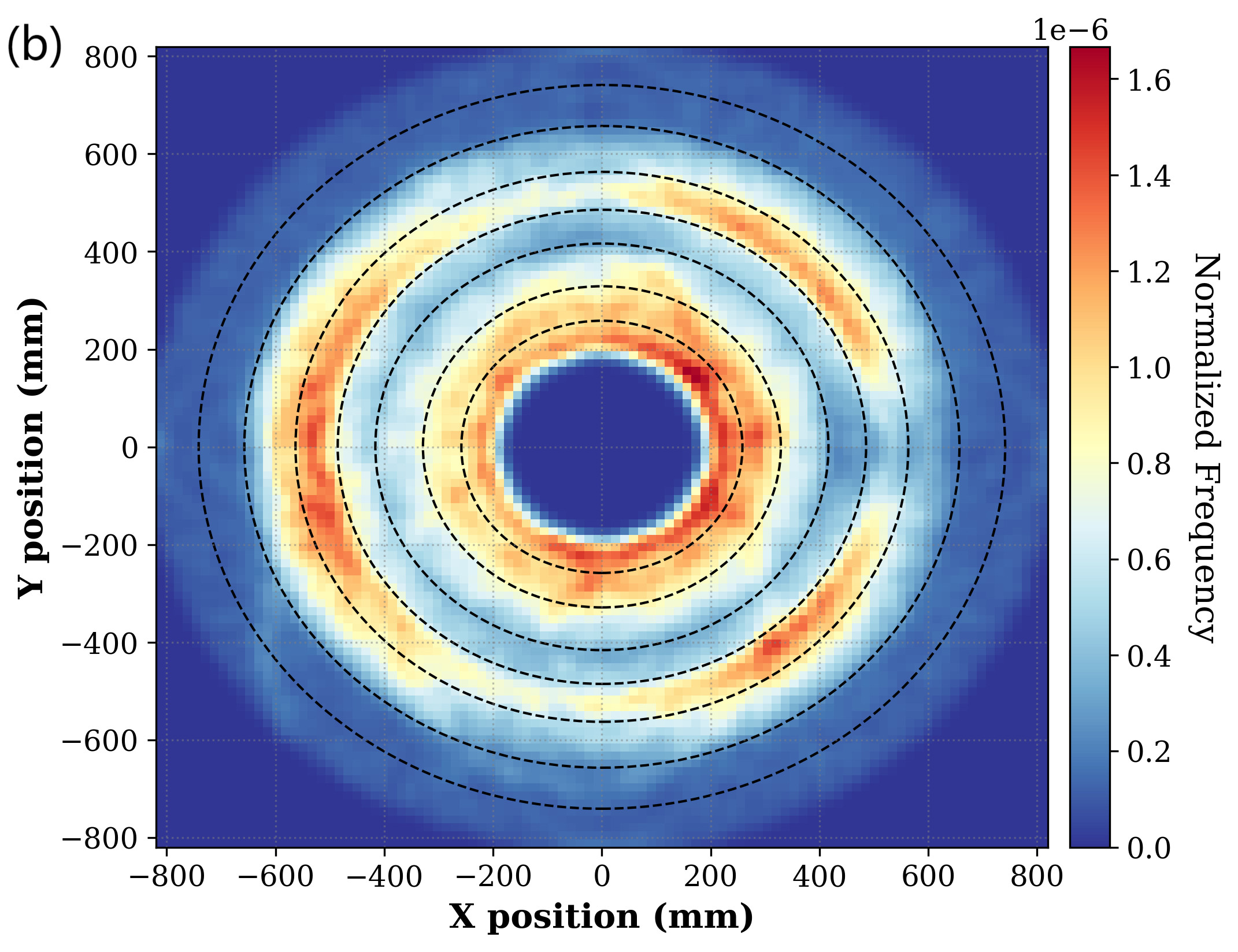}%
\hspace{0.01\textwidth}%
\includegraphics[width=.31\textwidth]{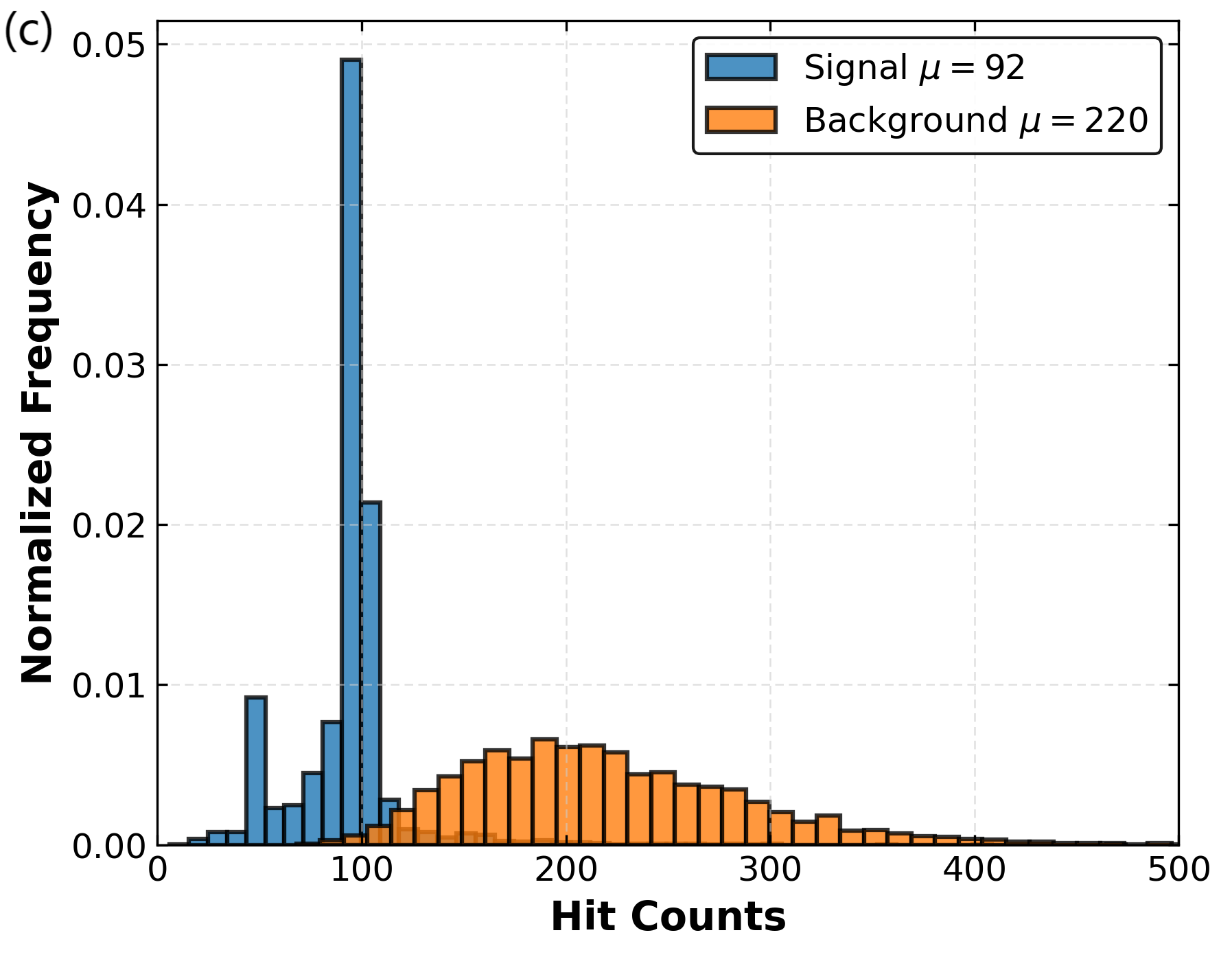}
\caption{(a): The distributions of cos$\theta$ versus $p_{T}$ for $\pi$ in $J/\psi \rightarrow \pi^0\pi^+\pi^- \rightarrow \gamma\gamma\pi^+\pi^-$ events. (b): Background spatial distribution in the $x-y$ plane of MDC with circular lines delimiting superlayer boundaries. (c): Signal (blue) and background (orange) hit count statistics.\label{fig:2}}
\end{figure}

The $\pi^{\pm}$ tracks from $J/\psi \rightarrow \pi^0\pi^+\pi^- \rightarrow \gamma\gamma\pi^+\pi^-$ decay, simulated via Geant4~\cite{geant4} within the OSCAR 2.6.0 framework, constitute the signal dataset with track momenta spanning a wide range e.g. [0.05, 1.5] GeV/c. The distribution of cos$\theta$ versus $p_{T}$ for $\pi^{\pm}$ is illustrated in figure~\ref{fig:2}a.
The spatial distribution of background generated by OSCAR 2.6.0 in the $x-y$ plane of MDC is shown in figure~\ref{fig:2}b, where the background exhibits a concentrated distribution in superlayers 0, 1, and 4. Figure~\ref{fig:2}c demonstrates the number of signal hits and noise in $J/\psi \rightarrow \pi^0\pi^+\pi^-$ Monte Carlo (MC) sample after mixing background, with a mean of about 92 for signal hits and a mean of about 220 background noises per event.

Two test datasets are established for algorithm validation:

Dataset 1: The $J/\psi \rightarrow \pi^0\pi^+\pi^- \rightarrow \gamma\gamma\pi^+\pi^-$ MC events mixed with the baseline background level (mean count around 220 hits/event)

Dataset 2: The $J/\psi \rightarrow \pi^0\pi^+\pi^- \rightarrow \gamma\gamma\pi^+\pi^-$ MC events mixed with twice the baseline background level (mean count around 440 hits/event), simulating the anticipated background escalation in the future.

\subsection{Signal hit selection and noise suppression}
\label{sec:4-2}

\begin{figure}[htbp]
\centering
\includegraphics[width=.4\textwidth]{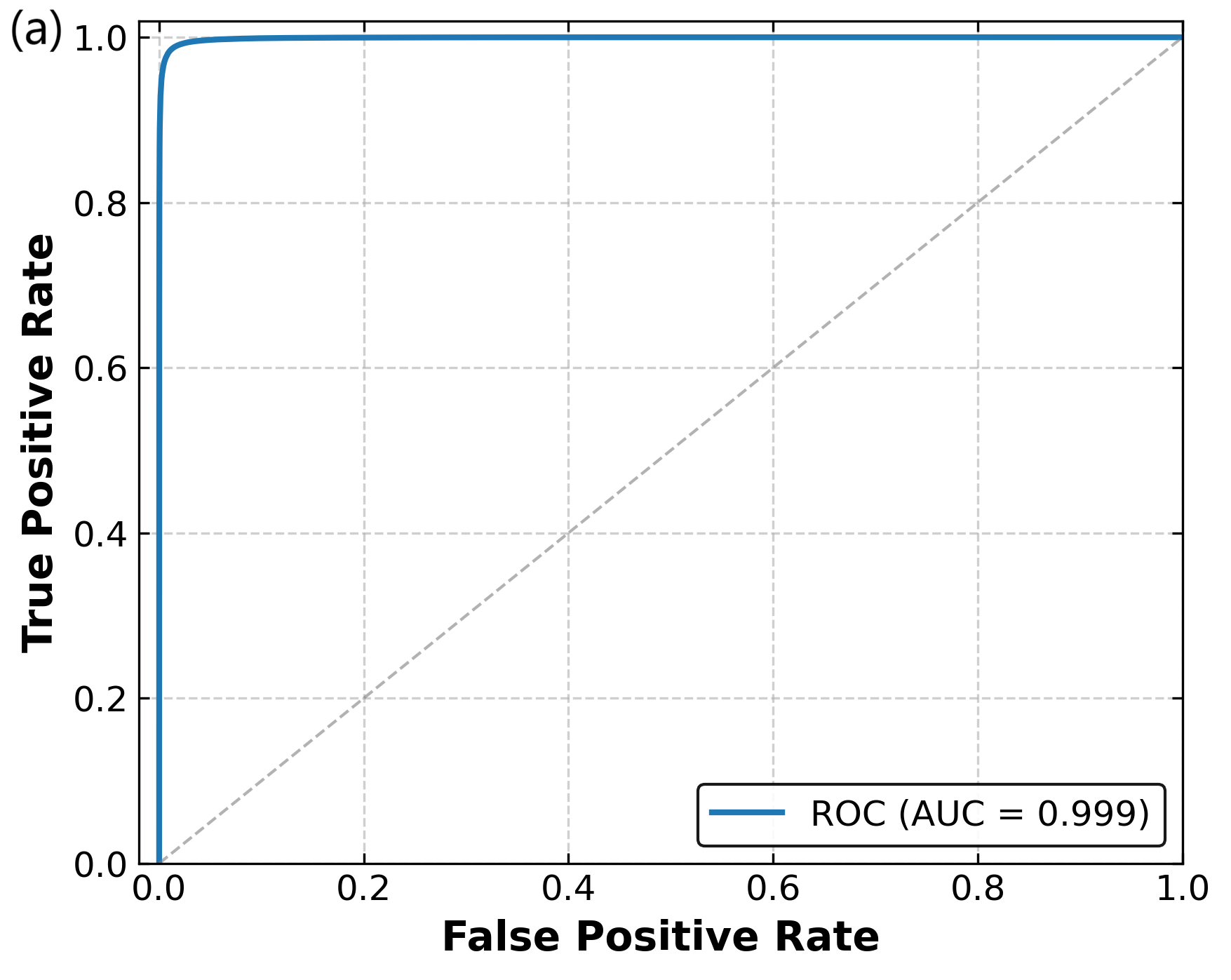}
\qquad
\includegraphics[width=.4\textwidth]{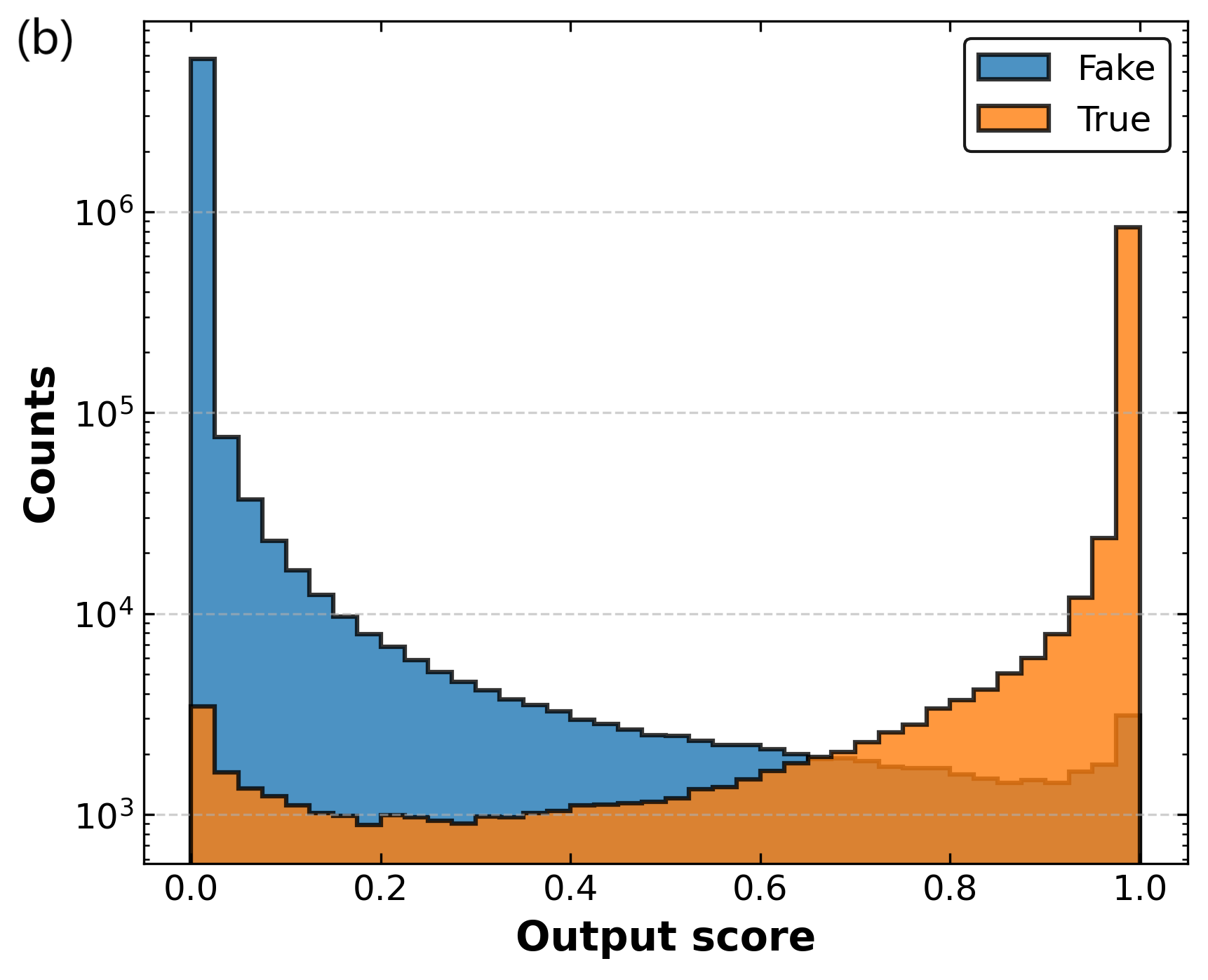}
\caption{Training results of the Graph Neural Network. (a): ROC with AUC = 0.999. (b): The distribution of the edge scores predicted by the GNN for true edges which connect the nodes that come from the same track, and for fake edges that do not.\label{fig:5}}
\end{figure}

The model's edge classification performance is presented in figure~\ref{fig:5}. Figure~\ref{fig:5}a shows the Receiver Operating Characteristic (ROC) curve, plotting the true positive rate against the false positive rate across classification thresholds, with a global AUC of 0.999. Figure~\ref{fig:5}b shows the edge score distribution, where true and fake track segments form distinct peaks at high and low scores, respectively. During validation, significant signal loss is observed in the innermost (layer 0) and outermost (layer 47) layers when applying the same threshold to all edges, due to their single adjacent layer configuration. As track fitting critically relies on signals from the inner layers, a tiered threshold strategy is adopted: no filtering for boundary layers (0 and 47), low-threshold filtering for inner layers, and high-threshold filtering for others.

Track reconstruction efficiency is the paramount metric for evaluating track reconstruction performance. As a preprocessing step for track reconstruction algorithms, noise suppression must preserve optimal reconstruction efficiency without compromising track quality, while minimizing fake rates.
These criteria collectively guide the optimization of parameters for the noise suppression algorithm.

Through parameter space scanning, the tiered threshold strategy configured with a low threshold of 0.1 in superlayer 0-1 and a high threshold of 0.5 in other layers achieves optimal track reconstruction performance. This configuration achieves a noise rejection rate of 86.8\% (ratio of excluded noise hits to total noise hits) while maintaining 98.2\% signal selection efficiency (ratio of retained true signals to total true signals).
These metrics demonstrate the algorithm's balance between noise suppression and signal preservation, fulfilling the requirements for high-precision trajectory reconstruction in high-background environments.

\begin{figure}[htbp]
\centering
\includegraphics[width=1.0\textwidth]{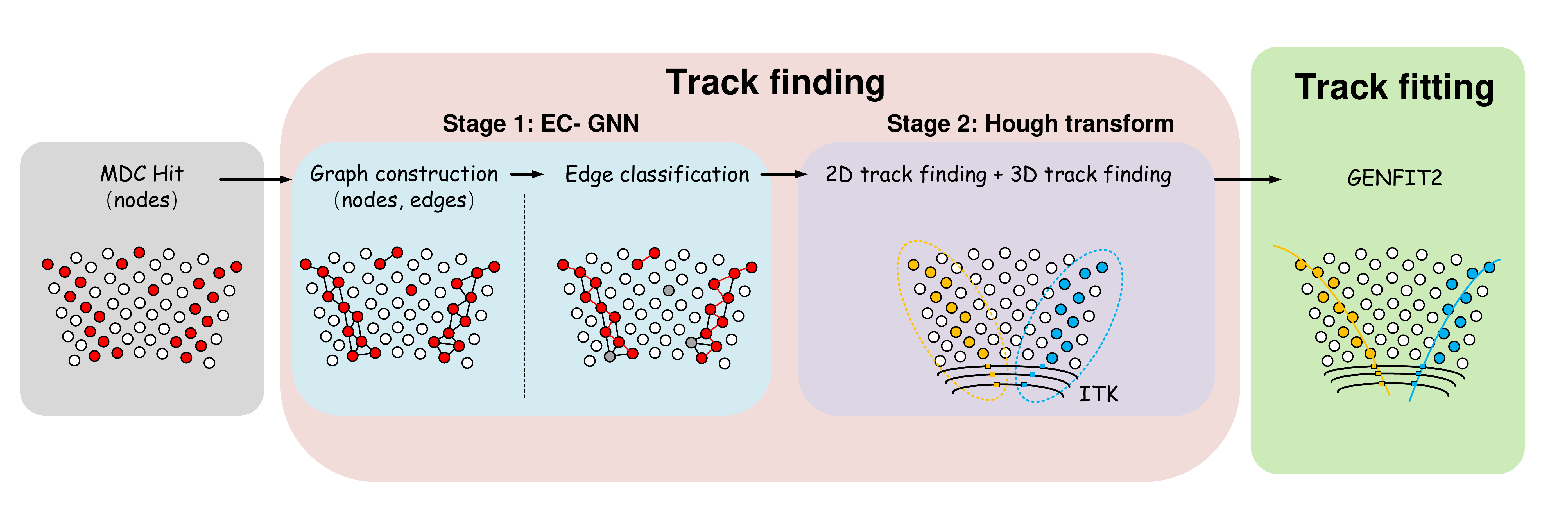}
\caption{Architecture of Graph Neural Network.\label{fig:flow}}
\end{figure}

\subsection{Performance of the combined GNF-Hough Algorithm}

The GNF Algorithm is added to the STCF tracking chain, prior to the Hough Transform track finding, to further evaluate its impact on the overall tracking performance, as shown in figure~\ref{fig:flow}. Metrics including track reconstruction efficiency, fake rate and spatial resolution are quantified across 200k simulated events. Track reconstruction efficiency is defined as the fraction of simulated particles with at least five measurable hits in the detector acceptance that are matched to reconstructed tracks. A track is matched to its primary particle if at least 80\% of its hits originate from that particle. Events containing reconstructed tracks unmatched to any primary particle are flagged as having fake tracks. The fake rate is then calculated as the ratio of events with fake tracks to the total number of events containing reconstructed tracks. 

\begin{figure}[htbp]
\centering
\includegraphics[width=.46\textwidth]{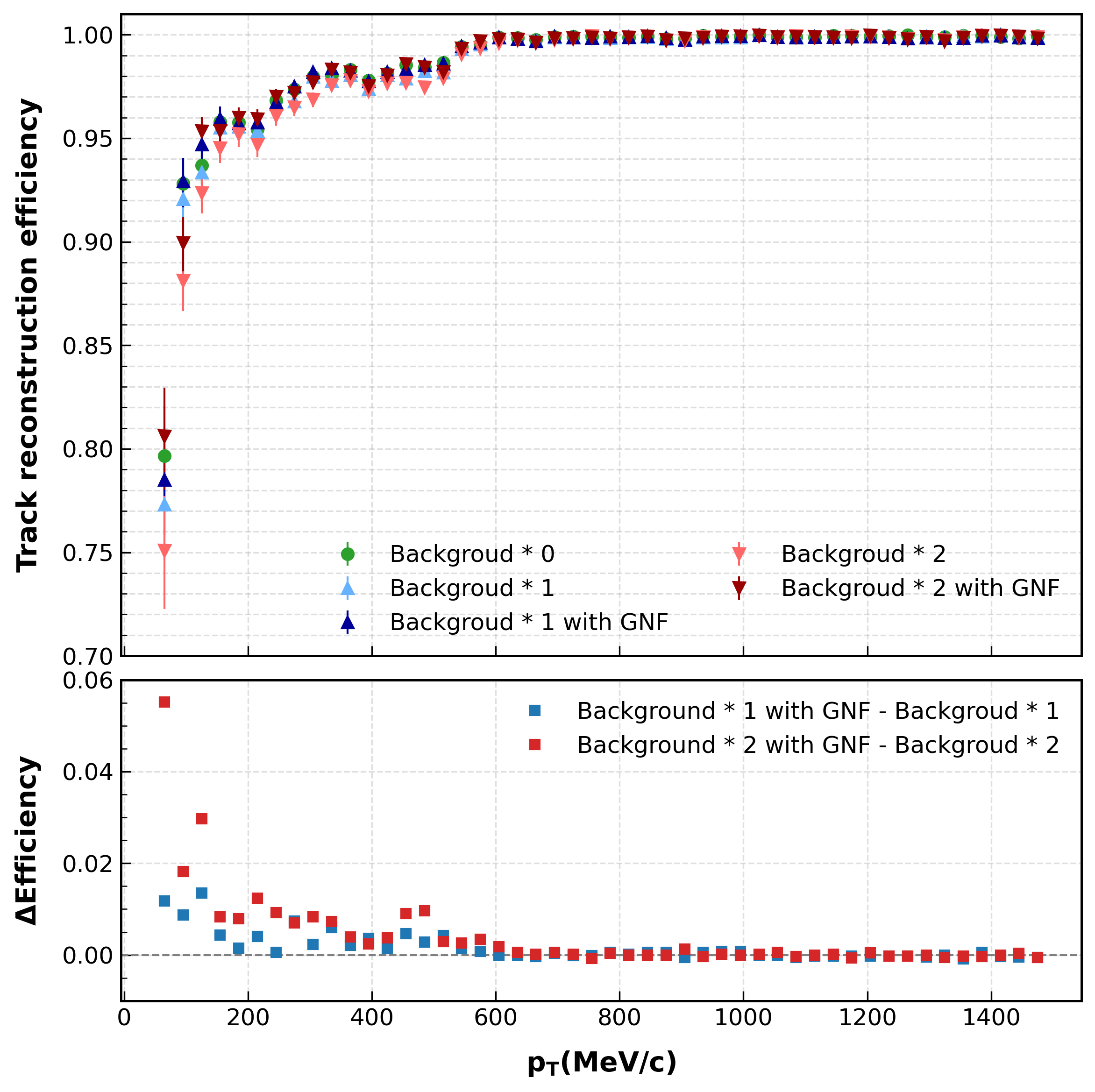}
\qquad
\includegraphics[width=.46\textwidth]{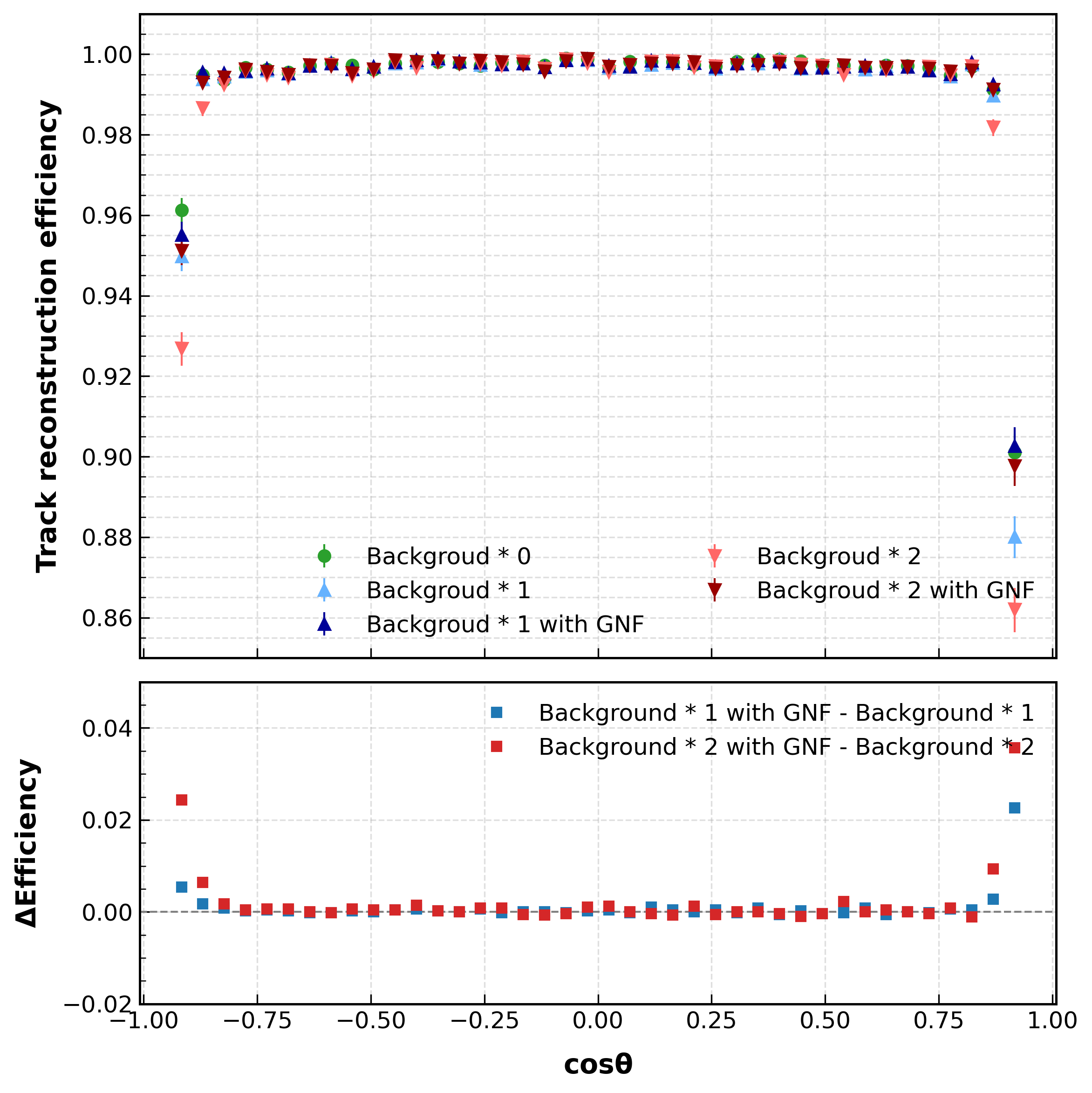}
\caption{Top: The track reconstruction efficiencies as functions of $p_{T}$ (left) and cos$\theta$ (right) for $\pi^{\pm}$ in $J/\psi \rightarrow \pi^0\pi^+\pi^- \rightarrow \gamma\gamma\pi^+\pi^-$ events. Green, light blue, and light red markers denote results without GNF under no background, standard background, and double background conditions, respectively. Deep blue and deep red markers indicate results with GNF at standard and double background levels, respectively. Bottom: Absolute efficiency gain achieved by GNF, blue and red squares represent improvements at standard background level and double background level conditions, respectively. \label{fig:6}}
\end{figure}

\begin{figure}[htbp]
\centering
\includegraphics[width=0.5\textwidth]{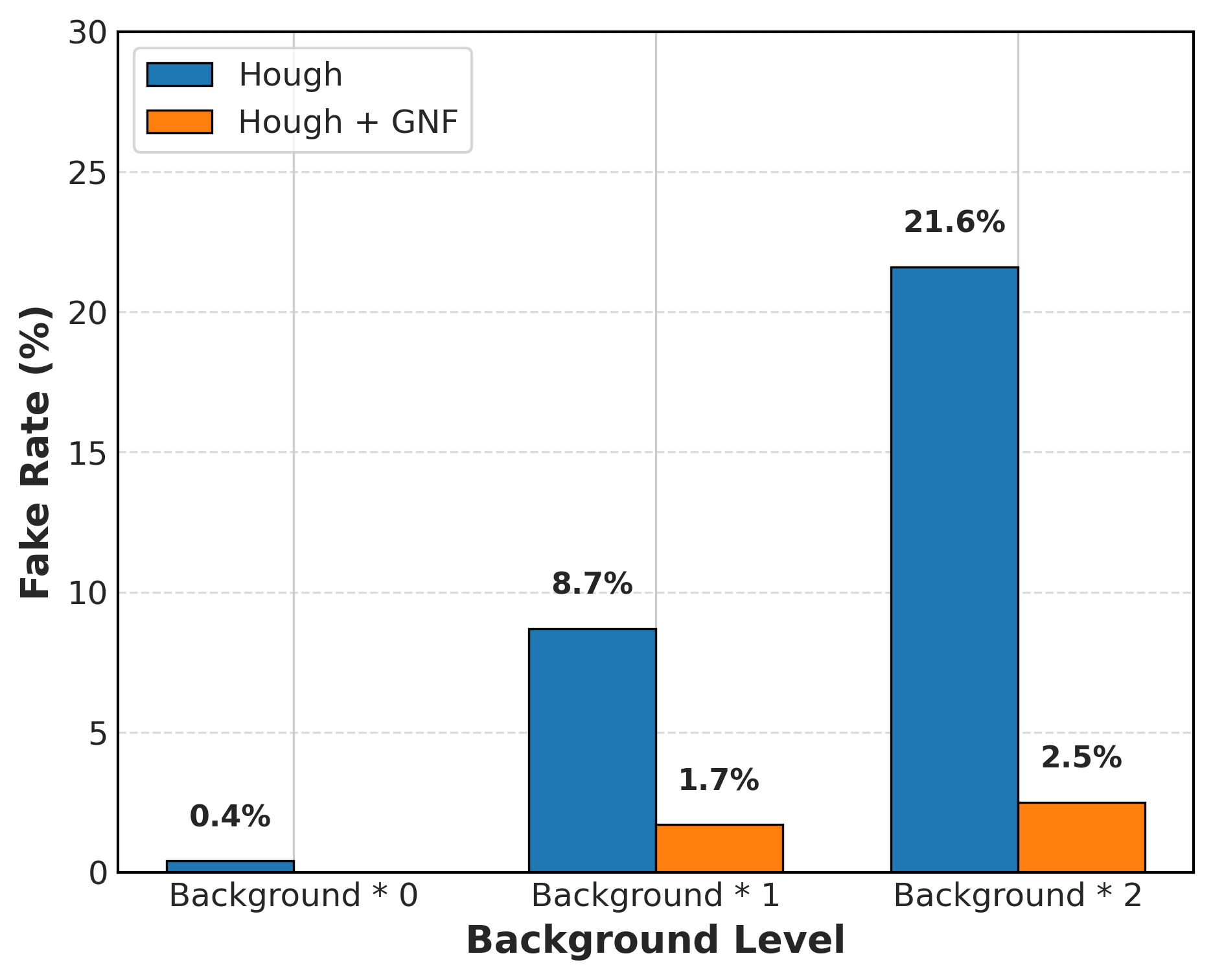}
\caption{The fake rate for $\pi^{\pm}$ in $J/\psi \rightarrow \pi^0\pi^+\pi^- \rightarrow \gamma\gamma\pi^+\pi^-$ events.\label{fig:10}}
\end{figure}

\begin{figure}[htbp]
\centering
\includegraphics[width=.32\textwidth]{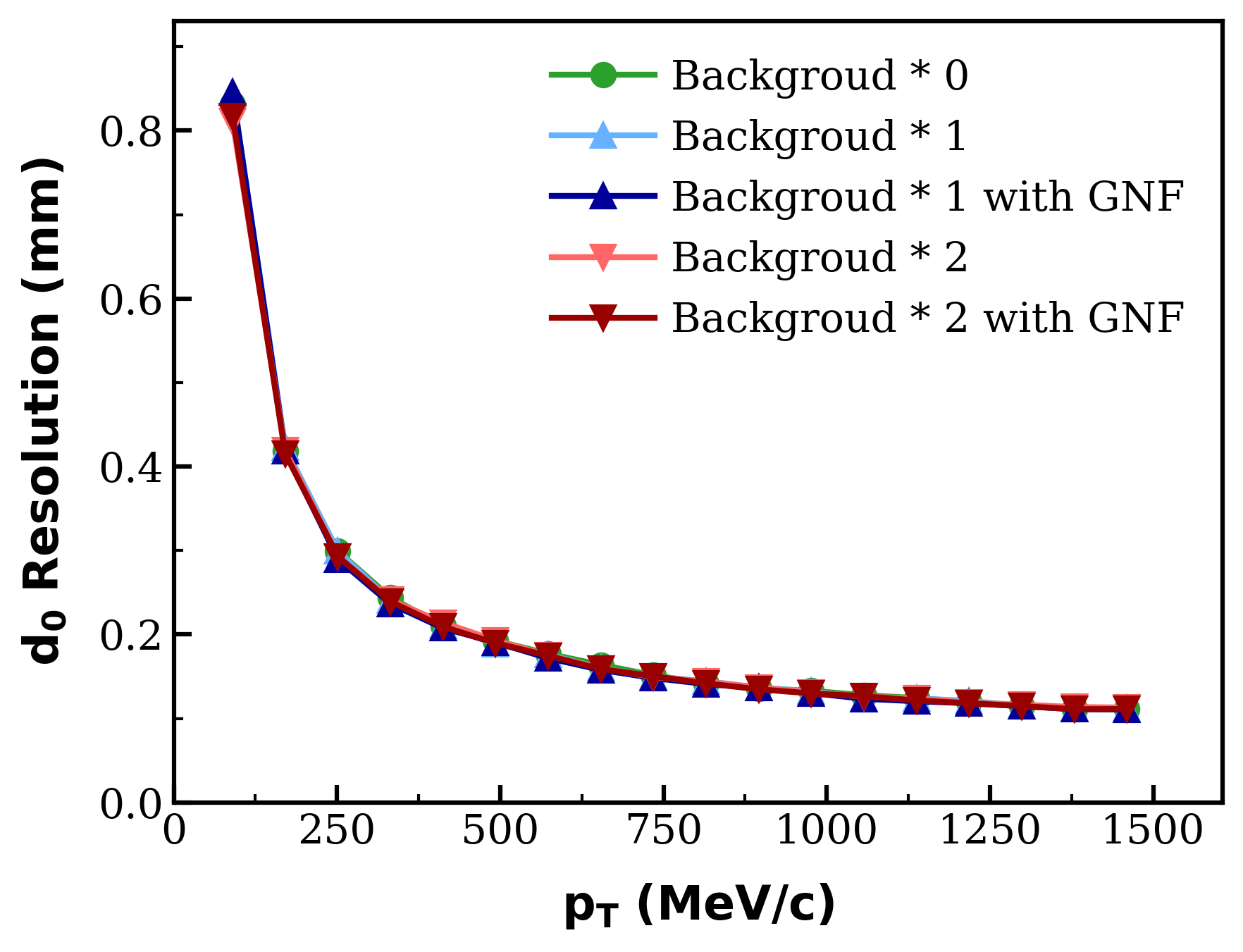}%
\hspace{0.01\textwidth}
\includegraphics[width=.325\textwidth]{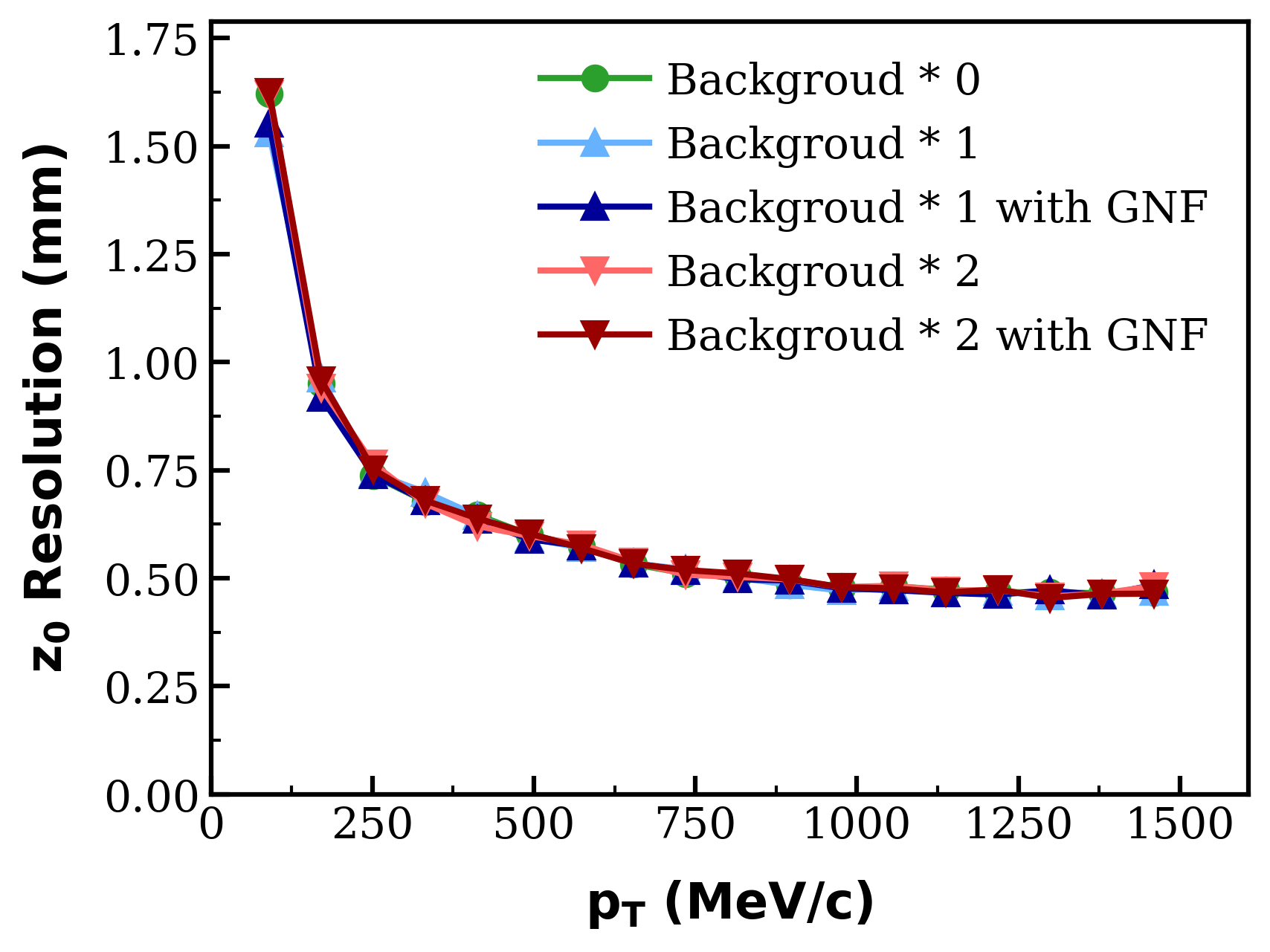}%
\hspace{0.01\textwidth}%
\includegraphics[width=.33\textwidth]{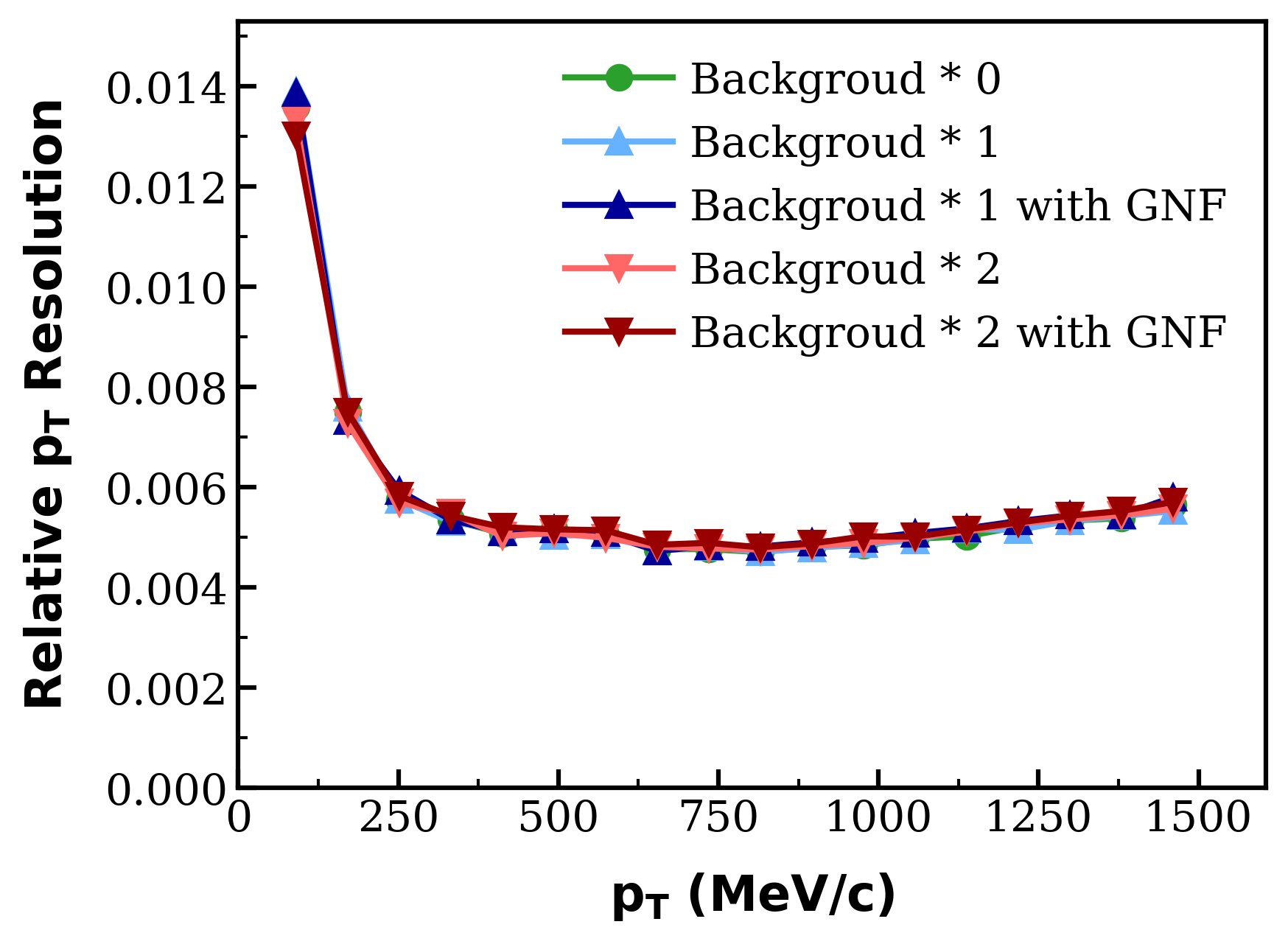}
\caption{The resolution of $d_{0}$ (left), $z_{0}$ (middle) and relative resolution of $p_{T}$ (right) for $\pi^{\pm}$ in $J/\psi \rightarrow \pi^0\pi^+\pi^- \rightarrow \gamma\gamma\pi^+\pi^-$ events. Green, light blue, and light red markers denote results without the GNF under no background, standard background, and double background conditions, respectively; deep blue and deep red markers indicate results with GNF at standard and double background levels, respectively.\label{fig:7}}
\end{figure}

Figure~\ref{fig:6} shows the track reconstruction efficiency versus transverse momentum and cos$\theta$ before and after applying GNF Algorithm for noise filtering. Significant performance recovery is observed at low transverse momentum ($p_{T}$ < 600 MeV/c) and large dip angles ( |cos$\theta$| > 0.8), where reconstruction accuracy approaches background-free levels post-denoising. Figure~\ref{fig:10} presents the algorithm's remarkable efficacy in suppressing fake tracks across different background conditions. The fake rate is reduced from 8.7\% to 1.7\% under standard background conditions --- an 80.5\% relative reduction. Even more significantly, at double the standard background level where the fake rate reaches 21.6\%, the algorithm reduces it to 2.5\%, achieving an 88.4\% reduction. Figure~\ref{fig:7} shows the resolution of impact track parameters $d_{0}$, $z_{0}$ and relative resolution of $p_{T}$ as a function of $p_{T}$. Here, $d_{0}$ is the signed distance from the point of closest approach (POCA) to the $z$-axis, and $z_{0}$ is the $z$ coordinate at the POCA. As $p_{T}$ increases, the resolution of these parameters initially decreases, reaches a minimum, and then deteriorates, with the relative resolution of $p_{T}$ exhibiting the most pronounced trend. This behavior arises because multiple Coulomb scattering dominates particle energy loss ($dE/dx$) at low transverse momenta. As $p_{T}$ increases, the contribution of multiple scattering diminishes, and the positional measurement precision becomes the dominant factor  determining the momentum resolution. At larger $p_{T}$, the reduced number of measurement points degrades the resolution of $p_{T}$ and other track parameters. As shown in figure~\ref{fig:7}, no significant difference was observed in the resolution of track parameters reconstructed via the Hough method across three background levels, while the track quality remained unaffected after noise filtering. By comparing these metrics with and without noise filtering, we can quantify the improvements achieved by the GNF Algorithm and demonstrate its beneficial impact on subsequent track processing tasks.

\section{Conclusion}
\label{sec:5}

We present a novel GNN-based noise filtering algorithm, the GNF algorithm, for the STCF drift chamber. The algorithm exploits the unique capabilities of GNNs to capture complex relationships within the data, enabling it to effectively distinguish between true and fake edges. Instead of pursuing an extremely high noise suppression rate, the algorithm adopts a tiered threshold strategy that achieves good reconstruction efficiency, extending GNN applications from edge classification to hit-level classification. Preliminary results demonstrate its capability to reduce noise while preserving particle track integrity, recovering the tracking efficiency loss caused by excessive noise in low-momentum track reconstruction, and effectively improving the purity of reconstructed tracks while ensuring track quality. Further optimizations are required to suppress fake tracks more aggressively, with the long-term goal of developing a universal noise-filtering framework adaptable to drift chambers across diverse experiments.

\acknowledgments
This work is supported by the National Natural Science Foundation of China (NSFC) under Contracts Nos.12025502, 12341504, 12175124, 12105158, 12188102.


\bibliographystyle{JHEP}
\bibliography{biblio}

\end{document}